\documentclass[lettersize, conference]{IEEEtran}
\usepackage{amsmath,amsfonts}
\usepackage{amssymb}
\usepackage{algorithmic}
\usepackage{algorithm}
\usepackage{array}
\usepackage[caption=false,font=normalsize,labelfont=sf,textfont=sf]{subfig}
\usepackage{textcomp}
\usepackage{stfloats}
\usepackage{url}
\usepackage{verbatim}
\usepackage{graphicx}
\usepackage{cite}
\usepackage{todonotes}
\usepackage{multirow} 
\usepackage{siunitx} 

\usepackage{bm} 
\usepackage{makecell}

\newcommand{\asinh}{\textrm{asinh}} 

\usepackage{threeparttable} 
\usepackage{booktabs} 

\begin{document}

\title{Throughput Maximization in Multi-Band Optical Networks with Column Generation}


\author{
\IEEEauthorblockN{Cao Chen\IEEEauthorrefmark{1}, Shilin Xiao\IEEEauthorrefmark{1}, Fen Zhou\IEEEauthorrefmark{2}, Massimo Tornatore\IEEEauthorrefmark{3}}
\IEEEauthorblockA{\IEEEauthorrefmark{1}State Key Laboratory of Advanced Optical Communication Systems and Networks, Shanghai Jiao Tong University, China}
\IEEEauthorblockA{\IEEEauthorrefmark{2} CERI-LIA, University of Avignon, France}
\IEEEauthorblockA{\IEEEauthorrefmark{3}Department of Electronics, Information and Bioengineering, Politecnico di Milano}
\IEEEauthorblockA{Emails:\{chen.cao, slxiao\}\{at\}sjtu.edu.cn, fen.zhou\{at\}univ-avignon.fr, massimo.tornatore\{at\}polimi.it}
}
 


\markboth{Journal of \LaTeX\ Class Files,~Vol.~X, No.~X, August~202X}%
{Shell \MakeLowercase{\textit{et al.}}: A Sample Article Using IEEEtran.cls for IEEE Journals}


\maketitle

\begin{abstract}
Multi-band transmission is a promising technical direction for spectrum and capacity expansion of existing optical networks.  
Due to the increase in the number of usable wavelengths in multi-band optical networks, the complexity of resource allocation problems becomes a major concern. 
Moreover, the transmission performance, spectrum width, and cost constraint across optical bands may be heterogeneous.
Assuming a worst-case transmission margin in U, L, and C-bands, this paper investigates the problem of throughput maximization in multi-band optical networks, including the optimization of route, wavelength, and band assignment.
We propose a low-complexity decomposition approach based on Column Generation (CG) to address the scalability issue faced by traditional methodologies. 
We numerically compare the results obtained by our CG-based approach to an integer linear programming model, confirming the near-optimal network throughput.
Our results also demonstrate the scalability of the CG-based approach when the number of wavelengths increases, with the computation time in the magnitude order of \SI{10}{\second} for cases varying from 75 to 1200 wavelength channels per link in a 14-node network.
Code of this publication is available at \url{github.com/cchen000/CG-Multi-Band}.
\end{abstract}


\section{Introduction}
\IEEEPARstart{I}{n} the past decade, traditional wavelength division multiplexing (WDM) optical networks characterized by fixed-grid frequency slots of \SI{50}{GHz} have moved toward flexible optical networks that benefit from a higher flexibility in assigning fiber spectrum resources, and which can now allocate variable-size frequency slots at granularities multiples of \SI{12.5}{GHz}. 
More recently, multi-band  (MB) transmission was proposed to exploit fiber spectrum resources beyond the C-band, e.g., L and U-bands\cite{SFNC20}.
For these bands, the spectrum width, transmission performance, and cost of devices are very different. These differences across different bands shall be carefully considered when performing the Route and Wavelength Assignment (RWA). 
Moreover, due to the increasing number of usable wavelengths in MB optical networks, the RWA complexity becomes a new matter of concern.

This paper investigates the problem of throughput maximization in MB networks that requires to solve the Route, Wavelength, and Band Assignment (RWBA).
We formulate RWBA as an integer linear programming (ILP) model and address its high complexity by proposing an approach based on column generation (CG). 
In this approach, a column corresponds to a \textit{colorless} configuration, describing how multiple routes are assigned to the same wavelength.
Each column is allowed to be repeatedly used across the whole spectrum.
This formulation had different names in prior studies regarding RWA, such as, among others, \textit{independent set}\cite{RaSi95} and \textit{routing configuration}\cite{LeLP00}. 
We choose here \textit{wavelength configuration}\cite{JaDa17} to highlight its strict relation to wavelength channel assignments.
After the first appearance of the concept of wavelength configuration in CG in \cite{LeLP00}, new and improved formulations \cite{JaMT09}, enhancements to the algorithm's scalability \cite{JaDa17}, and expansion to multicast routing \cite{ZhJA18} were studied.
In this paper, we extend the concept of wavelength configuration in CG to incorporate: \textit{(i)} flexible transceivers that can modulate the transmission parameters according to the quality of transmission (QoT) of lightpath, and (\textit{ii}) MB transmission, considering the heterogeneous requirements across optical bands, and addressing its inherent scalability issue.

\section{Network Model and Problem Statement}\label{sec: model}

This paper studies the RWBA problem considering a fixed-grid optical network. 
$G=(V,E)$ is used to denote an optical network with node set $V$ and link set $E$.
Length of a link $\ell\in E$ is characterized by $N_\ell$ spans.
Each link has a pair of fibers that support bidirectional traffic and an equal amount of spectrum resources.
We assume a spectrum width of \SI{15}{THz} per fiber including the U, L, and C-bands, with a bandwidth of \SI{5} THz per band\footnote{This paper assumes U+L+C bands due to the possible minimum fiber loss around its center frequency at 1570~nm \cite{ShNS22}.
It cannot be excluded that transmission devices for S-band will appear before those for U-band.}.
A lightpath $p$ denotes a simple path from source $s$ to destination $d$ on a particular wavelength. 
This wavelength should be consistent across all traversed links to maintain wavelength continuity. 
The carried transmission capacity is a function of the path length and the transmission margin requirement on its band (to be explained later).

Given a normalized traffic demand matrix $\hat{D}$, $\sum_{s,d}\hat{D}_{s,d}=1$,
we aim to evaluate the maximum throughput that a network can achieve, i.e., the total data assigned for the demands. 
We need to allocate transmission capacities for each node pair using adequate lightpaths, subject to the following constraints:
\begin{itemize}
    \item \textit{Wavelength conflict constraint}: no two lightpaths can use the same wavelength on the same link;
    \item \textit{Wavelength limitation constraint}: available wavelengths should not exceed the spectrum resources. 
    More attention is required when modeling the RWBA problem, since different bands may hold different spectrum widths;
    \item \textit{Demand-capacity constraint}: traffic demand between two nodes should not exceed the total transmission capacity between them, so that we can assess the network performance in a non-blocking state;
    \item \textit{Cost constraint}: we express the possible cost constraint from network operators by limiting the number of transceivers in a network, or lightpath count equivalently. 
\end{itemize}

The transmission capacity of a lightpath varies with the modulation format based on the signal-to-noise ratio (SNR). 
We consider the modulation formats \{PM-BPSK, PM-QPSK, PM-8QAM, PM-16QAM, PM-32QAM, PM-64QAM, PM-128QAM, PM-256QAM\}\footnote{These formats are with net spectral efficiencies of \{1.6, 3.1, 4.7, 6.3, 7.8, 9.4, 10.9, 12.5\} in \SI[]{}{\bit\per\second\Hz} and minimum required SNRs of \{3.7, 6.7, 10.8, 13.2, 16.2, 19, 21.8, 24.7\} in dB scale\cite{SaVI19}.}. 
Once the SNR of a lightpath exceeds the minimum required SNR, any modulation format from the candidate list can be chosen. The transmission capacity is given as follows, 
\begin{align}
\label{eq: Capacity}
    C = S_m \cdot R_{s}
\end{align}
with the spectral efficiency of the $m^\text{th}$ modulation format $S_m$, and transceiver's baud-rate $R_s$.

For the QoT evaluation, we selected a physical layer model that can capture the effect of band-dependent transmission impairments, while being simple enough to permit the scalable QoT computation required for a networking study. 
We adopt the approach in \cite{SeKB17} that computes the nonlinear interference based on the conventional incoherent Gaussian Noise model. 
This approach first computes the \textit{effective attenuation coefficient} $\alpha_{\text{eff},w}$ and \textit{effective length} $L_{\text{eff},w}$ for a wavelength channel $w$ on a span with fiber attenuation coefficient $\alpha_{w}$ \cite{ShNS22} and span length $L_s$. 
The optical signals experience the inter-channel stimulated Raman scattering (ISRS) effect that can be approximated by a linear function with slope $C_R$ and cutoff frequency $f_R$.
The power loss and tilt after each span are assumed to be fully compensated. 
For a fully-loaded reference spectrum with a center frequency $f_c$, a channel spacing $B_{ch}$, total bandwidth $B_t$, wavelength set
$\{w\in \mathbb{Z}\left|1\leq w\leq W\right.\}$, a uniform power level $P_0$ per channel, and total power $P_t$,  the SNR for the $w^\text{th}$ channel  {can then be calculated} as follows,
\begin{align}
\label{eq: SNR}
SNR_w&= P_0/\left(P_{\text{ASE}, w}+ \eta_w P_0^3\right)\\
    P_{\text{ASE},w} &=\exp\left( \alpha_{\text{eff},w} L_s\right)  \cdot \text{NF} \cdot R_s \cdot h \cdot (f_c+f_w)\\
    \eta_w & \approx \frac{8}{27} \gamma_w^2 \alpha_{\text{eff},w} L_{\text{eff},w}^2 \frac{\asinh \left( \pi^2 |\beta_2 | B_t^2 /(2\alpha_{\text{eff},w}) \right)}{\pi|\beta_2 | R_s^2}
\end{align}
with noise figure $\text{NF}$,  Planck's constant $h$, nonlinear coefficient $\gamma_w$\cite{ShNS22}, channel frequency {$f_c+f_w$}, and group velocity dispersion $\beta_2$.
We assume an ideal rectangular Nyquist-WDM spectrum for each channel and the channel spacing equal to the baud-rate, and ignore the impact of transceiver noise and filtering effect. 

\begin{figure}
    \centering
    \includegraphics[width=0.36\textwidth]{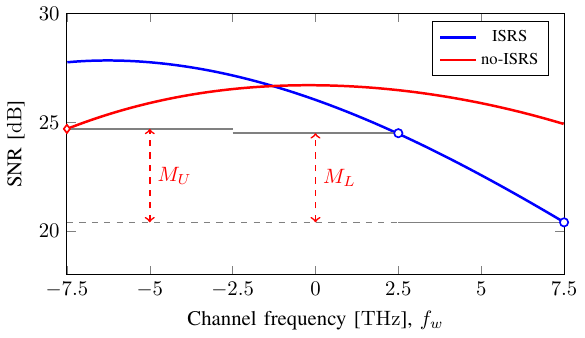}
    \caption{
     Illustration of the SNR {on different channel frequencies $f_w$} after the first span assuming the ISRS (blue) and no ISRS (red), respectively. 
	 The transmission margins $M_U$ and $M_L$ are obtained by differentiating the worst-case SNR for that optical band {(solid gray line, RWBA)} and for all bands {(dashed gray line, RWA)}. 
	 This paper computes the upper bound of SNRs on different optical bands by simply considering two states of ISRSs.
    }
    \label{fig: distributionSNRonULC}
\end{figure}

Next, we describe the key assumption used to set the margins in the three bands. 
We transform the wavelength-dependent SNR into band-dependent SNR by choosing the minimum SNR on that optical band. 
As such, we evaluate the transmission performance on each optical band under full-load worst-case assumptions, regardless of the actual deployments of optical amplifiers and lightpaths. 
Given the minimum SNR after the first span as $SNR_{b}$, the SNR after $N$ spans is given as follows,
\begin{align}
\label{eq: SNR_computation}
SNR_{b,\SI{}{\dB}}^{(N)} = SNR_{b,\SI{}{\dB}}^{(1)} - 10\log_{10} N - M_{b,\SI{}{\dB}}
\end{align}
where $M_{b,\SI{}{\dB}}$ is a margin preserved for the $b^\text{th}$ band. 
Based on (\ref{eq: SNR}), we plot the SNR for different channel frequencies in Fig.~\ref{fig: distributionSNRonULC}, where we also report the minimum SNR of three bands. 
The power spectral density of \SI[per-mode=symbol]{14}{\mu W \per \GHz } and single-mode fiber are assumed. 
In RWBA, we consider $24.8$, $24.5$, and \SI{20.4}{\dB}.
Instead, when solving RWA (i.e., when we do not differentiate the performance of the three bands), the SNR margins are considered to be $4.4$, $4.1$, and \SI{0}{\dB}, respectively, assuming the lightpaths on all bands work with a flat SNR of about \SI{20.4}{\dB} ($\approx120$), so that we can neglect the band allocation while always ensuring successful transmission.

\section{Integer Linear Programming Model}\label{sec: ILP model}

The following subsection \ref{subsec: ILP} formulates the problem of RWBA with throughput maximization as an ILP model.  
Note that the ILP is based on path formulation, where we precalculate $K$ routes via Yen's algorithm\cite{Yen71}.
As such we can easily pre-estimate the SNR of each lightpath via (\ref{eq: SNR_computation}) and choose the maximum capacity from the highest-order modulation format {compatible with that SNR}.
Then, in subsection \ref{sec: small example}, we use an example on a 4-node network to clarify some of the issues of the ILP, and point out a promising solution.
 
\subsection{ILP Model}\label{subsec: ILP}

\textbf{Parameters}
\begin{itemize}
\item $A$: maximum number of available transceivers.
\item $w \in \{1, 2, \cdots, W\}$: wavelength index. 
\item $\hat{D}_{s,d}$: normalized traffic demand profile, $\sum_{s,d} \hat{D}_{s,d} = 1$.
\item $p_{s,d,k}$: a path ID of the $k^\text{th}$ shortest path between $(s,d)$.
\item $\beta_{s,d,k,\ell}$: equals 1 if  $p_{s,d,k}$ uses fiber link $\ell$, 0 otherwise.
\item $C_{s,d,k,w}$: max. transmission capacity for $p_{s,d,k}$ on $w$ assuming the highest  order modulation format. 
\end{itemize}

\textbf{Variables}
\begin{itemize}
\item $\delta_{s,d,k,w}$:  equals 1 if the $p_{s,d,k}$ adopts the  $w^\text{th}$ wavelength, 0 otherwise.
\item $D_{s,d}$: traffic demand between $(s,d)$.
\item  {$T_{s,d}$: transmission capacity between $(s,d)$.}
\item $TH$: network throughput.
\end{itemize}
\allowdisplaybreaks
\begin{subequations}
\label{model: P1_SB}
\begin{align}
\!  {\max\limits_{\delta_{s,d,k,w}\in \mathbb{B}}}  & TH
 \quad \boldsymbol{} && \nonumber  \\
 \label{eq: con1_inthroughput_normalized}
\textbf{\textit{s.t.~~~}} &  D_{s,d}=TH\cdot \hat{D}_{s,d} \quad\quad && \forall (s, d) \\
\label{eq: con1_inthroughput_b}
 &  D_{s,d} \leq  T_{s,d}  \quad \quad \quad \quad &&\forall (s, d) \\
\label{eq: conAddi_inthroughput_DT}
&  T_{s,d} = \sum\nolimits_{k}\sum\nolimits_{w} \delta_{s,d,k,w}   C_{s,d,k,w} &&\forall (s, d) \\
\label{eq: con3_inthroughput_spectrumSlot_b}
& \sum\nolimits_{s} \sum\nolimits_{d\neq s} \sum\nolimits_{k}  \delta_{s,d,k,w}  \beta_{s,d,k,\ell} \leq 1  &&\forall \ell, w \\
\label{eq: con3_cost}
& \sum\nolimits_{s} \sum\nolimits_{d\neq s} \sum\nolimits_{k}\sum\nolimits_{w}  \delta_{s,d,k,w} \leq A 
\end{align}
\end{subequations}
The objective is to maximize the network throughput. 
Constraints (\ref{eq: con1_inthroughput_normalized}) ensure the traffic {demands} distribution is compliant with the given {parameters}.  Constraints (\ref{eq: con1_inthroughput_b}) are the demand-capacity constraints. 
Constraints (\ref{eq: conAddi_inthroughput_DT}) compute the transmission capacities of the lightpaths connecting $(s,d)$. 
Constraints (\ref{eq: con3_inthroughput_spectrumSlot_b}) avoid wavelength clash for all wavelengths on each link. 
Constraint (\ref{eq: con3_cost}) is the cost constraint used for limiting the transceiver count. 

\subsection{A Small Example}\label{sec: small example}

\begin{figure}[!tbp]
    \centering
    \includegraphics[width=0.36\textwidth]{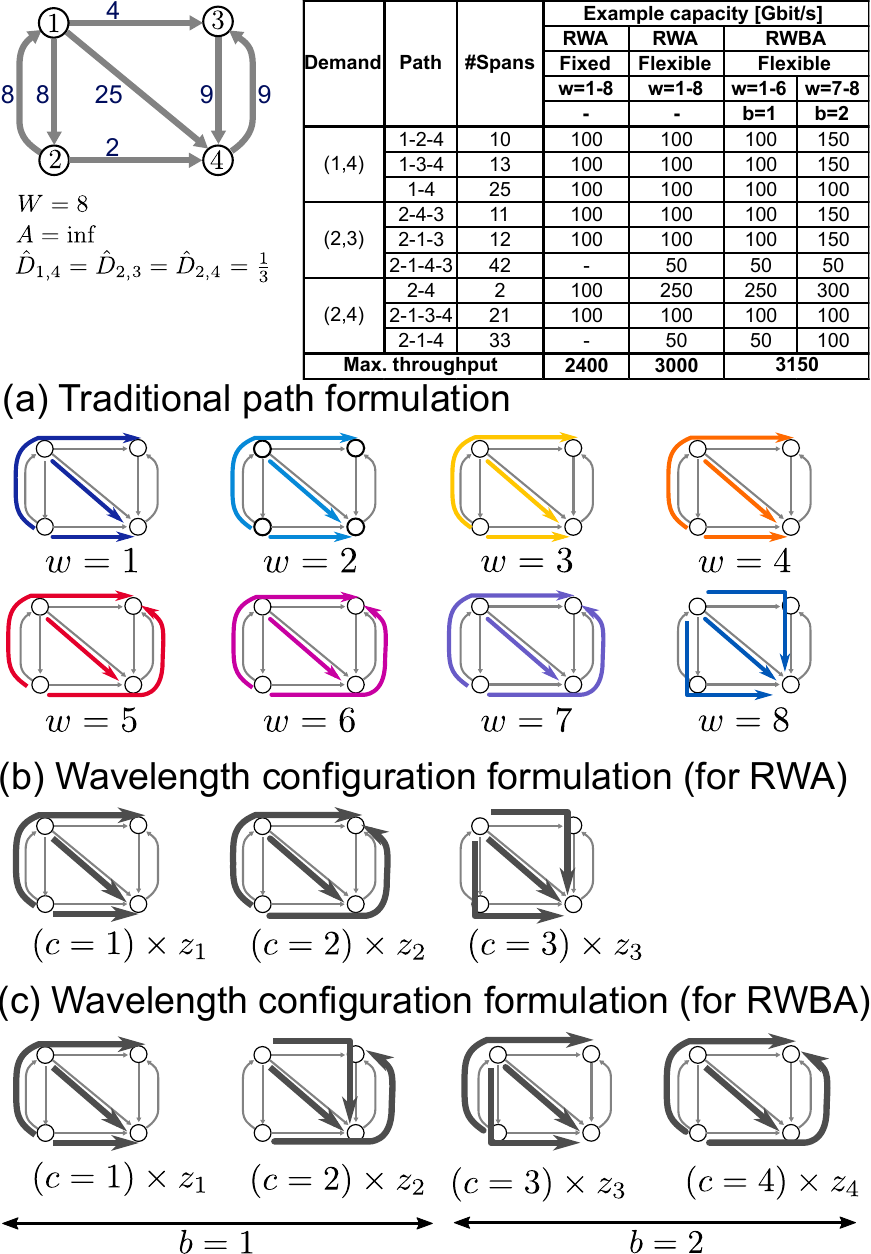}
    \caption{
    Illustration of a 4-node network with the span length on each fiber and with a table on the right illustrating the transmission capacities for \num{9} paths on \num{8} wavelengths, $w\in\{1,...,8\}$, or on \num{2} bands, $b\in\{1,2\}$. 
	(a), (b), and (c) show the respective assignment based on path formulation and wavelength configuration formulations, $c\in\{1,2,3,4\}$.}
    \label{fig: CG_illustration}
\end{figure}

For the above model, scalability becomes a concern in the case of a large number of wavelengths.
Let us consider the 4-node network in Fig.~\ref{fig: CG_illustration} and discuss the variables required to optimize \num{3} demands $(1,4)$, $(2,3)$, and $(2,4)$ when given \num{3} candidate routes and \num{8} wavelengths.
In the previously-discussed path-based formulation, the main variable is boolean $\delta_{s,d,k,w}$ and at least $3\times 3 \times 8$ variables should be used,
\begin{align}
\max_{\delta_{s,d,k,w}\in \mathbb{B}} \left\{ TH \left| TH\leq \min\left\{ \frac{ T_{1,4}}{\hat{D}_{1,4}}, \frac{T_{2,3}}{\hat{D}_{2,3}},
\frac{T_{2,4}}{\hat{D}_{2,4}}
\right\}, \text{(\ref{eq: conAddi_inthroughput_DT})-(\ref{eq: con3_cost})}\right.\right\}
\end{align}
where $\mathbb{B}$ is \{0, 1\} set. 
A possible assignment can refer to Fig.~\ref{fig: CG_illustration}(a). 
For notation simplicity we may denote a path directly by its traversed nodes when expressing $C_{s,d,k,w}$ and drop $w$ if no capacity difference across wavelengths. 
Thus, $T_{1,4} = 
\delta_{1,4,3,1}C_{p_{14},1} + \delta_{1,4,3,2}C_{p_{14},2} + \delta_{1,4,3,3}C_{p_{14},3} + \cdots = 8C_{p_{14}} + C_{p_{134}} + C_{p_{124}}$, $T_{2,3} = 7 C_{p_{213}} + 3 C_{p_{243}}$, and $T_{2,4} = 5C_{p_{24}}$.
For this formulation, the number of variables increases linearly with the number of wavelengths. 
As the solution space grows exponentially with the number of variables, excessive computing time might soon be needed to search for the optimal solution.

Inspired by the route similarity of the wavelengths 1, 2, 3, and 4, we propose to apply the concept of ``wavelength configuration'' to solve the throughput maximization problem such that a given routing configuration on a wavelength is captured as one single variable.
The first graph in Fig.~\ref{fig: CG_illustration}(b) shows this configuration, marked in gray to express that it is currently colorless.
Note that such a similar idea has been used in prior studies \cite{JaDa17, LeLP00}, but prior studies only considered lightpaths with fixed capacity that can always achieve successful transmission for arbitrary lengths.
Instead, we herein study the wavelength configurations enabled by flexible transceivers with a distance-adaptive transmission capacity. 

As an example, in Fig.~\ref{fig: CG_illustration}(b) and Fig.~\ref{fig: CG_illustration}(c), we show some possible configurations and compare the maximal $TH$ between fixed and flexible transmission capacity and between RWA and RWBA.
The transmission capacity $T_{s,d}$ in  Fig.~\ref{fig: CG_illustration}(b) is now given as follows,  $\left[T_{1,4}, T_{2,3}, T_{2,4}\right] =z_1  [C_{p_{14}}, C_{p_{213}}, C_{p_{24}}] + z_2 [C_{p_{14}}, C_{p_{213}} + C_{p_{243}}, 0] + z_3   [C_{p_{14}}+C_{p_{124}}+C_{p_{134}}, 0, 0]$, where $z_c$ denotes the number of times a configuration is used.
For the fixed transmission capacity, the maximal $TH$ of \SI[per-mode=symbol]{2400}{\giga\bit\per\second} can be achieved with the optimal solution $z_\text{RWA}^*=(8,0,0)$.
For the flexible transmission capacity, the maximal $TH$ of \SI[per-mode=repeated-symbol]{3000}{\giga\bit\per \second} is achieved with $z_\text{RWA}^*=(4,3,1)$.
The network throughput can be further increased if we account for RWBA that preserves a lower SNR margin on the $2^{\text{nd}}$ band.
In this case, four new configurations are required, as shown in Fig.~\ref{fig: CG_illustration}(c). 
The maximal $TH$ of \SI[per-mode=symbol]{3150}{\giga\bit\per\second} is now achieved with the optimal solution $z_\text{RWBA}^*=(5,1,1,1)$. 
Note that the optimization variables required in the above cases have changed from boolean $\delta_{s,d,k,w}$ to integer $z_c$, reducing the memory consumption from \num{72} to \num{3} or \num{4}.
 
This example shows the advantage of wavelength configuration in reducing the optimization variables, how flexible transmission capacities can be incorporated within the wavelength configuration concept, and how wavelength configuration effectively lends itself to application for design of MB optical networks.  
However, the number of potential configurations can also grow exponentially in real network scenarios.  
So, in the next section, to more efficiently identify and generate high-quality configurations, we present our proposed CG approach.

\section{Column Generation Approach}\label{sec: CGs}

We begin by introducing CG decomposition in RWA to understand the iterative process for generating a new column, including the restricted master problem (RMP) and pricing problem.
We then turn our attention to RWBA to deal with different optical bands.
The CG approaches for RWA and RWBA are presented in subsections \ref{sec: CG for RWA} and \ref{sec: CG for RWBA}, respectively.  
Algorithm~\ref{alg: alg1} shows the outline of CG.
In line~\textbf{\ref{alg: initial}}, we initialize the wavelength configuration set by a sequential loading algorithm {based on} $k$-Shortest-Path ($k$SP) and First-Fit (FF).
Up to $|V|(|V|-1)$ configurations, or $W$ configurations if $W\leq |V|(|V|-1)$, are filled. 
As $k$SP-FF is also used for benchmarking, we provide the details in Sec.~\ref{sec: simulation_setup}.

\begin{algorithm}[!ht]
\caption{Column Generation Approach}\label{alg: alg1}
\begin{algorithmic}[1]
\STATE Initialize wavelength configurations $\Omega_\text{init}=\{c_1, c_2, ...\}$\label{alg: initial}
\STATE $c_\text{new} \longleftarrow \emptyset, \Omega\longleftarrow\Omega_\text{init}$
\STATE \textbf{Do} \label{alg: begin_it}
\STATE \hspace{0.5cm} $\Omega \longleftarrow \Omega \cup \{ c_\text{new} \}$
\STATE \hspace{0.5cm} Solve  RMP$(\Omega)$ with {\textit{relaxing}} variables, $z_c\in \mathbb{R}_{\geq 0}$
\STATE \hspace{0.5cm} Choose a column $c_\text{new}$ via pricing problem{(s)}\label{alg: PP} 
\STATE \textbf{While} 
Objective in line~\textbf{\ref{alg: PP}} is  {positive}\label{alg: end_it} 
\STATE Solve  RMP($\Omega$)  with \textit{integer} variables, $z_c\in \mathbb{Z}_{\geq 0}$ \label{alg: RMP_integer}
\STATE  {Assign a} wavelength  {to each} configuration \label{alg: RMP_color}
\end{algorithmic}
\label{alg1}
\end{algorithm}

\subsection{CG  for RWA}\label{sec: CG for RWA}

\textbf{Parameters}
\begin{itemize} 
\item $c\in \Omega$ : wavelength configuration ID. 
\item $C_{s,d,k}$ : transmission capacity of path $p_{s,d,k}$.
\item $\delta_{s,d,k,c}$ : equals 1 if path $p_{s,d,k}$ is used by the configuration $c$, 0 otherwise.
\item $T_{s,d,c}$   : total transmission capacities for node pair $(s,d)$  in configuration  $c$, $T_{s,d,c}  = \sum_{k} C_{s,d,k} \cdot \delta_{s,d,k,c}$. 
\item $a_{c}$         : number of transceivers in a configuration, $a_{c} = \sum_{s}\sum_{d\neq s} \sum_k \delta_{s,d,k,c}$. 
\end{itemize}

\textbf{Variables}
\begin{itemize}
\item $z_c$:  number of times configuration $c$ is used.
\end{itemize}

\subsubsection{RMP} 
\begin{subequations}
\label{model: MP_RWA}
\begin{align}
\notag
\!\!\!\! \max_{z_c \geq 0} & ~TH \quad\boldsymbol{}    \\
\!\!\!\! \textbf{\textit{s.t.~}}  &\hat{D}_{s,d}\cdot TH +  \sum\nolimits_{c\in\Omega} (-T_{s,d,c}) z_c  \leq 0\,  \forall (s,d)&[\sigma_{s,d}] \label{eq: demand-capacity constraint}   
 \\
  &  \sum\nolimits_{c\in\Omega} a_c  z_c     \leq A   &[\sigma_{A}] \label{eq: limited lightpath constraints} \\
 &  \sum\nolimits_{c\in\Omega}  z_c  \leq W    &[\sigma_{W}] \label{eq: wavelength limit constraints} 
\end{align}
\end{subequations}
where $\sigma_{s,d}$, $\sigma_A$, and $\sigma_{W}$ are the dual variables for demand-capacity constraints (\ref{eq: demand-capacity constraint}), cost constraint (\ref{eq: limited lightpath constraints}), and wavelength limitation constraint (\ref{eq: wavelength limit constraints}). 

\subsubsection{Pricing Problem} According to the dual of RMP, we can derive the pricing problem as follows,
\begin{subequations}
\label{model: Pricing_RWA}
\begin{align}
 \notag
\max\limits_{\delta_{s,d,k,c'}\in \mathbb{B}}  &  \sum\nolimits_{s}\sum\nolimits_{d\neq s} \sigma_{s,d} \left(+ T_{s,d,c'}\right) - a_{c'} \sigma_{A}  - \sigma_{W}   \\
\label{eq: pp_nonoverlapping}
\textbf{\textit{s.t.}}   & \sum\nolimits_{s}\sum\nolimits_{d\neq s} 
 \sum\nolimits_{k} \delta_{s,d,k,c'} \beta_{s,d,k,\ell} \leq 1 \quad \forall \ell \\
\label{eq: pp_def1}
&  T_{s,d,c'} = \sum\nolimits_{k} C_{s,d,k}  \delta_{s,d,k,c'}  \quad \forall (s,d)
\\
\label{eq: pp_def2}
&  a_{c'} = \sum\nolimits_{s}\sum\nolimits_{d\neq s} \sum\nolimits_{k} \delta_{s,d,k,c'} 
\end{align}
\end{subequations}
where $c'$ is a configuration under design, with multiple routes denoted by $\delta_{s,d,k,c'}$. 
Constraints (\ref{eq: pp_nonoverlapping}) ensure no wavelength clash on this configuration. 
Constraints (\ref{eq: pp_def1}) and (\ref{eq: pp_def2}) calculate the variables in the objective function.
 The configuration is deemed promising if the objective can be greater than 0.

Regarding the first two terms of the objective, this pricing problem can also be expressed as a maximum-weight independent set problem, where each node denotes a physical path with weight $\sigma_{s,d} C_{s,d,k}-\sigma_{A}$. 
Rather than dealing with the time-consuming ILP model, we propose a heuristic algorithm, which
finds the node set with the maximum weight by sequentially selecting a node based on the descending order of node weight, until the weight of the next node is 0 or negative.
Meanwhile, we skip those nodes that create wavelength clash with the already chosen nodes.

Next, we describe the utilization of the columns. 
Recalling that the variable denoting lightpath usage must be an integer, we compute such a solution by rerunning the RMP and restricting the variable $z_c$ to be an integer.
Then, we end CG with a simple wavelength packing strategy that groups the same wavelength configuration together, sorts them by the configuration times, and assigns contiguous wavelengths.

\subsection{CG  for RWBA} \label{sec: CG for RWBA}
For RWBA, a wavelength configuration has to specify an optical band so that the specific requirements of different bands can be considered. 
This is fulfilled by adding two parameters $\psi_{c,b}$ and $C_{s,d,k,b}$ to note whether the configuration $c$ belongs to optical band $b$ and the transmission capacity matrix of each band.

\textbf{Additional Parameters}
\begin{itemize}
\item  $b\in \mathcal{B}$ : optical band ID, $\mathcal{B}$=\{U, L, C\}.
\item $W_b$  : available wavelengths on the $b^{\text{th}}$ optical band.
\item $\psi_{c,b}$ : equals 1 if wavelength configuration $c$ is used for the $b^{\text{th}}$ optical band, 0 otherwise.
\item $C_{s,d,k,b}$ : transmission capacity of $p_{s,d,k}$ on the $b^\text{th}$ band. 
\end{itemize}

\subsubsection{RMP}
\begin{subequations}
\label{model: MP}
\begin{align}
\notag
 \max_{z_c\geq 0} \quad &TH  \quad\boldsymbol{\texttt{}}     \\
 \textbf{\textit{s.t.~~~}}  & (\text{\ref{eq: demand-capacity constraint}}), (\text{\ref{eq: limited lightpath constraints}}) \notag\\
 & \sum\nolimits_c \psi_{c,b}  z_c\leq W_b \quad\quad \forall b\in \mathcal{B} & [\sigma_{W,b}] \label{eq: wavelength limit constraints_mul}
\end{align}
\end{subequations}
Constraints (\ref{eq: wavelength limit constraints_mul}) restrict the wavelength usage on respective optical bands.

\subsubsection{Pricing Problem} 
\begin{subequations}
\label{model: PP in RWBA}
\begin{align}
\notag
 \max_{
 \begin{smallmatrix}
  \delta_{s,d,k,c'}&\in \mathbb{B}\\
  \psi_{c',b} &\in \mathbb{B}
 \end{smallmatrix}
 }   &  \sum_s\sum_{d\neq s} \sigma_{s,d} \left( + T_{s,d,c'}\right) - a_{c'} \sigma_{A} - \sum_b \psi_{c',b} \sigma_{W,b}   \\
 \textbf{\textit{s.t.~}}  &  
\text{(\ref{eq: pp_nonoverlapping})},\text{(\ref{eq: pp_def2})} \notag  \\
 &  T_{s,d,c'} = \sum\nolimits_{k,b} C_{s,d,k,b} \, \psi_{c',b} \, \delta_{s,d,k,c'}   \forall (s,d)\\
  & \sum\nolimits_b \psi_{c',b}=1 
\end{align}
\end{subequations}

We consider the following algorithm for the above model.
Given that an optical band is unique to a wavelength configuration, we decompose this pricing problem into $|\mathcal{B}|$ subproblems, each assuming the configuration is aligned to a specific optical band. 
We then apply the same algorithm used for the model (\ref{model: Pricing_RWA}) and choose the column with the most positive reduced cost.

\section{Illustrative Numerical Results}\label{sec: simulations}

We now evaluate the computational efficiency and network performance of ILP, CG, and two benchmark algorithms. 
We choose RWA to evaluate the impact of a large number of wavelengths in MB optical networks. 
Then, we compare the RWBA against RWA in order to evaluate the impact of the flexible capacity and MB transmission.

\subsection{Simulation Setup}\label{sec: simulation_setup}
We consider two real networks for Deutsche Telekom AG (DT), 9-node and 17-link and 14-node and 23-link noted as DT9 and DT14, respectively\cite{ViIS19}
with uniform traffic demand distribution among the node pairs i.e., $\hat{D}_{s,d}=\frac{1}{|V|(|V|-1)}$.

\subsubsection{Number of Wavelengths}
As our aim is to compare the computational efficiency for different numbers of wavelengths, we vary the number of wavelength channels by choosing different channel spacings depending on varying the baud of the deployed transceivers\footnote{Although we could also achieve the same goal by expanding wavelengths on a broader spectrum {beyond U+L+C bands}, this approach requires tuning the physical layer settings, including the margin and transmission capacity matrix. 
This adaptation involves a more complex physical layer modeling that can be considered as future work.}. 
Hence, the number of wavelength channels varies from 30 to 75, 100, 150, 300, 600, 1000, 1200, and 1500, based on transceivers operating at 500, 200, 150, 100, 50, 25, 15, 12.5, and 10 GBaud, respectively. 
The transmission margins of $4.4$, $4.1$, and \SI{0}{\dB}, and other physical parameter settings are shown in Sec.~\ref{sec: model}.

\subsubsection{Benchmark Algorithms}
We consider two sequential loading algorithms $k$SP-FF and FF-$k$SP with $10$ candidate routes. 
$k$SP-FF establishes a lightpath by searching all viable wavelengths for a given route, and turns to the next possible route, whereas FF-$k$SP searches all possible routes on a given wavelength and turns to the next possible wavelength\cite{ViIS19, CZLX22}. 
For the two algorithms, we uniformly load the network with a fixed-rate demand that equals the minimum positive transmission capacity of all lightpaths, so that at most one lightpath is established for a new demand. 
The network throughput and lightpaths are recorded every time a demand is loaded.
Blocking occurs if a demand cannot be fulfilled by any lightpath. 
When the first blocking occurs, the algorithm stops and returns the network throughput.

For the ILP in (\ref{model: P1_SB}), the relative integer solution gap is set to 5\% and the maximum simulation time is \SI{1}{\hour}. 
For the RMP with integer constraint, 
these parameters change to  1\% and \SI{10}{\second} in order to attain optimal performance within a reasonable computation time.
Note that, the typical solving time for the relaxed RMP is around \SI{2}{\ms \per round} for DT9 and \SI{50}{\ms \per round} for DT14 by using Gurobi 9.0 API and Matlab R2017b on a computer with 8 GB RAM and a 4-core Intel i5-6300HQ.

\subsection{Performance Comparison in RWA}
Fig.~\ref{fig: ex_time} shows the computational time and network throughput as the number of wavelength channels increases. 
The results of the two benchmarks and CG are averaged over \num{1000} independent simulations.
The points at which 100, 25, and 12.5~GBaud transceiver is assumed are marked by diamond, pentagon, and circle symbols, respectively.

\subsubsection{Computational Efficiency}
\begin{figure}[!tbp]
    \centering
    \includegraphics[width=0.45\textwidth]{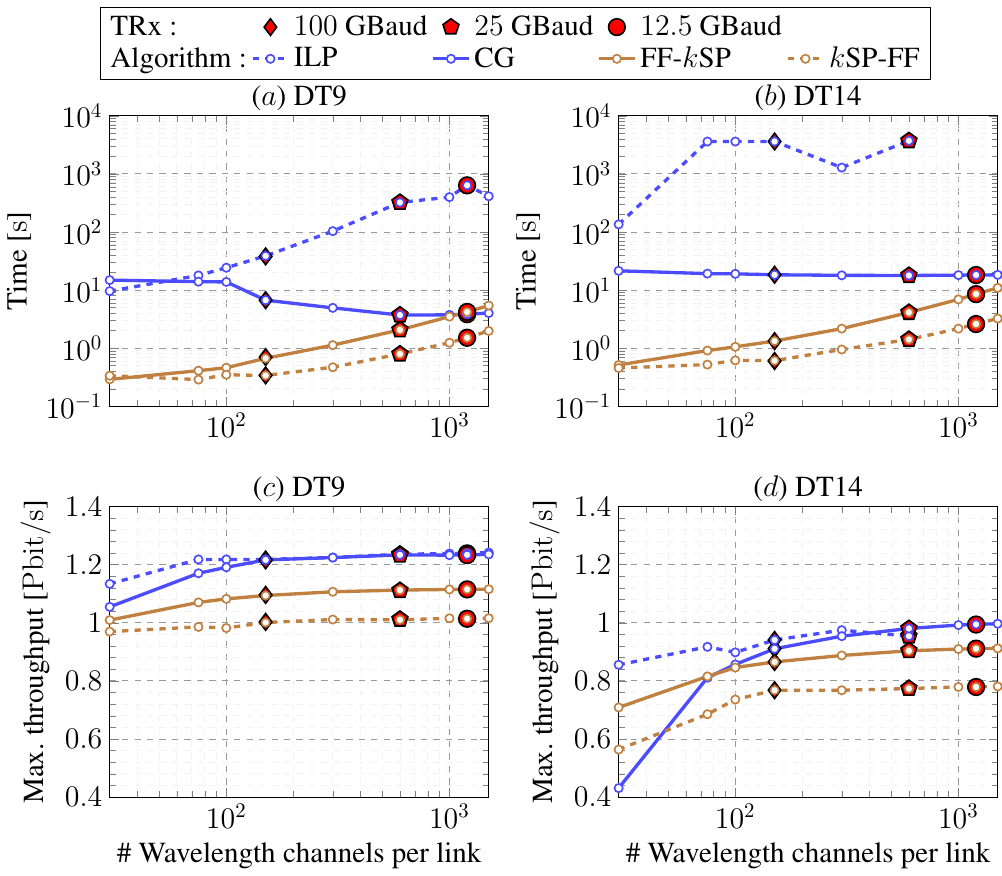}
    \caption{
    Computation time (top row) and network throughput (bottom row) for ILP, CG, FF-$k$SP, and $k$SP-FF algorithms, for DT9 and DT14 networks, and for variable numbers of wavelength channels in \SI{15}{\THz}.}
    \label{fig: ex_time}
\end{figure}

In Figs.~\ref{fig: ex_time}(a) and (b), the computation time for the two benchmark algorithms and ILP increases as the number of wavelengths increases.
This is caused by the algorithms' strategy that carefully searches the exact route and wavelength for each lightpath. 
This requires more time as the number of usable wavelengths increases. 
In addition, the ILP can not be established on DT14 that has wavelengths more than \num{1000} due to memory limitations.
 
The CG, unlike traditional methodologies, shows strong scalability as the number of wavelengths increases, maintaining a stable time consumption in the magnitude order of \SI{10}{\second}.
This approach focuses on the parallel design of each configuration rather than a specific lightpath, thus not sensitive to the number of wavelengths.
Counterintuitively, the computation time slightly decreases as the number of wavelengths increases. 
For CG, the time-consuming module mainly comes from the slow convergence in lines \textbf{\ref{alg: begin_it}}-\textbf{\ref{alg: end_it}} of Algorithm~\ref{alg: alg1}, and finding the integer solution in line \textbf{\ref{alg: RMP_integer}}.
The time saving comes from the latter. 
As the number of wavelengths increases, the integrality gap of the whole problem decreases. 
This means that the last RMP with integer constraint could be approximated as relaxed linear programming (LP) that has a fractional number of wavelengths so that the advantage of high computation efficiency inherent to LP is obtained.

\subsubsection{Max. Throughput}
Regarding the Figs.~\ref{fig: ex_time}(c) and (d), all algorithms exhibit an increase in the network throughput as the number of wavelengths increases. 
This is well understood because a more fine-grained spectrum is allowed as we vary the channel spacing and baud-rate of each transceiver.
Note that the CG easily outperforms the two benchmarks in most cases and is even comparable to the ILP model.
A small solution gap between the ILP model and CG exists, meaning that the current CG approach is near-optimal instead of optimal. 
We also note an exception in Fig.~\ref{fig: ex_time}(d) at the point with 30 wavelengths where the CG is worse than the others.
Most columns in such a case are used only once, meaning that the RMP in finding the integer solution is back into a complex 0-1 knapsack problem, which needs more time instead of the \SI{10}{\second} in our setting. 
Increasing the time, altering the initial configurations, and introducing a branch-and-bound algorithm can improve the performance.
Nonetheless, this exception does not represent a big issue for CG, as the typical number of wavelengths exceeds 80, e.g., CBand transmission on a 50~GHz grid.

\subsection{Performance Comparison in RWBA}
 
\begin{figure}[!tbp]
    \centering
    \includegraphics[width=0.40\textwidth]{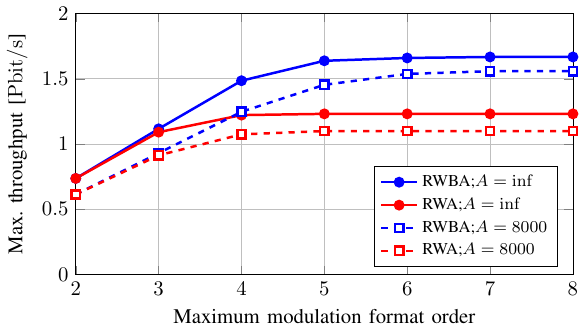}
    \caption{Maximum throughput of the DT9 using a 25~GBaud transceiver with variable modulation formats compared for the RWA (red) and RWBA (blue), and for the unlimited (filled circles) and limited transceivers (unfilled squares).}
    \label{fig: RWBAvsRWA}
\end{figure}

Next, we turn to RWBA and compare its performance against RWA.
We choose DT9 and 25~GBaud transceiver as an example, and report the results in Fig.~\ref{fig: RWBAvsRWA} with filled circle symbol.
Solving RWBA allows higher throughput, as the worst-case SNR in the respective optical bands is typically greater than the value assumed by RWA.
Besides, allowing a larger number of modulation formats increases the network throughput by 67\% for RWA and 125\% for RWBA.

As network operators are always concerned about the cost, we also consider the achievable throughput under a cost constraint represented by a limited number of transceivers, 8000 as an example.
The results are reported with unfilled square symbol in Fig.~\ref{fig: RWBAvsRWA}. 
Adding a cost constraint decreases the achievable throughput, but the results also confirm the benefit of the adapted modulation format and the heterogeneous MB transmission requirements.

\section{Conclusion}\label{sec: conclusion}

We have proposed and studied the network throughput maximization problem in MB fixed-grid optical networks with distance-adaptive modulation formats, including the optimization of route, wavelength, and band assignment.
An ILP model and a low-complexity CG-based approach were presented.
Beginning with the traditional RWA problem, we have validated the scalable performance of CG compared to the ILP model as the number of wavelengths increases.
Illustrative numerical results, tested on realistic network topologies with numbers of wavelengths per link varying from 75 to 1200, show that the CG is able to find the near-optimal solution in the magnitude of \SI{10}{\second}. 
This enables the CG approach to scale to a greater network, whereas the ILP model is not always scalable.
When turning to the novel RWBA problem, we observe a larger network throughput compared to the results obtained by solving the RWA problem, since the modulation formats can be adapted to suit the transmission margin requirements of these optical bands.
Results with adapted modulation format show a network throughput increase of 67\% and 125\% in the RWA and the RWBA problems for the Germany DT9 network, respectively. 
This also highlights the significant potential of combining the CG approach with flexible transceivers. 



\textbf{Acknowledgment}: Massimo Tornatore acknowledges the National Science Foundation (Grant No. 2226042) for funding.

\bibliographystyle{IEEEtran}
\bibliography{references}
\end{document}